\documentclass{article}

\usepackage{url}
\usepackage{algorithm,algpseudocode}
\usepackage{amsmath} 
\usepackage{comment}

\title{The Ritva Blockchain: Enabling Confidential Transactions at Scale}
\author{Henri Aare, Peter Vitols \\ 
Crystal Technology Research}
\date{May, 2020}

\begin{document}

\maketitle

\def\codename{\textsf{RVChain}}

\begin{abstract}
\textit{The distributed ledger technology has been widely hailed as the breakthrough technology. It has realised a great number of application scenarios, and improved workflow of many domains. Nonetheless, there remain a few major concerns in adopting and deploying the distributed ledger technology at scale. In this white paper, we tackle two of them, namely the throughput scalability and confidentiality protection for transactions. We learn from the existing body of research, and build a scale-out blockchain platform that champions privacy called \codename. \codename\ takes advantage of trusted execution environment to offer confidentiality protection for transactions, and scale the throughput of the network in proportion with the number of network participants by supporting parallel shadow chains. }

\end{abstract}

\section{The current state of affair}

The distributed ledger technology (DLT), or commonly referred to as blockchain technology, provides a transparent and secure manner to record transactions and manage assets. The transparency comes from the fact that the ledger can be made public, and thus it is subject to universal audit. The security is guaranteed thanks to cryptographic mechanisms that make transactions, once recorded on the blockchain, are irreversible. These very properties enable DLT to attract a significant amount of attention. The technology has gained enormous traction since the birth of Bitcoin (BTC)~\cite{btc_origin}. The two most popular blockchain networks at this time of writing are Bitcoin and Ethereum~\cite{eth_origin} networks. These two blockchains together manage hundreds of billions of dollars in assets. As impressive as this number appears, it only touches a very small fraction of the amount of assets currently circulated in the market. There remains a gigantic room for improvement, expansion, and adoption. We pay our primary attention to privacy and processing throughput of the transactions. 

The first hindrance that deters a wide adoption of blockchain technology in our everyday life is its limited scalability~\cite{hybster, rapidchain, tendermint, algorand}. While conventional centralized brokers such as VISA and PayPal can process thousands of transactions per second, the Bitcoin network only supports 5-8 transactions per second~\cite{pow_sec_vs_tps}. Ethereum, which is proclaimed to be the general-purpose smart-contract enabled blockchain platform, is only able to handle approximately 20 transactions per second~\cite{ohie, hybster}. Another factor that is worth considering is the non-finality and confirmation latency. A transaction on the Bitcoin network takes up to an hour to be reliably confirmed (i.e., the transaction’s relevant parties are confident that it will not be reversed), whereas that number for the Ethereum network is about 10 minutes~\cite{eth}. These weaknesses pose too much of a burden on any large-scale financial service. 

There is yet another concern that Bitcoin and Ethereum networks admit, which is its lack of privacy/confidentiality protections for transactions. All transactions and/or data posted on these blockchains are not only visible to their relevant parties, but also to the public. That is, they cannot safely store or compute on sensitive data (e.g., clinical record, financial transactions)~\cite{ekiden}. While both Bitcoin and Ethereum networks adopt pseudonyms so as to offer a certain level of privacy protections, a large body of research has shown that de-anonymizing such pseudonyms is feasible~\cite{sasson2014zerocash, li2017survey}.

\section{Bridging the gap with the \codename}
We at Ritva set out to mitigate the aforementioned current state of affair by developing a performative general-purpose blockchain platform that supports confidential transactions, called \codename. Abstractly, The \codename~ takes advantage of the recent development in computer hardware, in particular CPUs that are capable of provisioning Trusted Execution Environment (TEE)~\cite{sgx, sgx_guide, TPM, sealed_glass_proof, sgx_remote_attest, trustzone,sanctum}. Another technical feature that enables the \codename~ to operate at scale is its capability to support parallel shadow chains whose transactions can be totally ordered~\cite{ohie}.

By incorporating TEE, \codename~ effectively simplifies the threat model it has to deal with to crash fault tolerance~\cite{paxos,livepaxos, consensus_survey,quorum}, to which a number of performative consensus protocols have been studied. To further scale the transaction throughput of the platform alongside with its network size, \codename~ allows virtually unlimited number of shadow chains (subject only to the network size) to complement one another. Network participants (or miners) are assigned to a chain uniformly at random, and leverage crash fault tolerance consensus protocol to establish a total ordering of transactions on that chain. \codename~ then relies on a simple solution to establish a global order for transactions across the parallel shadow chains, thereby attaining consistency. The preliminary design of the \codename~ currently supports approximately 2500 transactions per second for the network consisting of 100 participants. Its theoretical foundation allows the transaction to be processed at the network speed. That is, the only limit to the performance of the \codename~ is the speed at which it receives transactions. With 5G technology on the horizon, the capacity is virtually limitless.

The other advantage of the TEE is its isolated execution~\cite{sgx}. More specifically, the TEE provisions a protected address space. Code and data running and being processed in this protected address space is inaccessible to any unauthorised processes, even the privileged ones such as the Operating System or Hypervisor. When there is a need to write the sensitive data off the protected address space to the secondary  memory, the TEE architecture transparently encrypts the sensitive data with cryptographic keys only available to the processor. \codename~ makes use of this isolated execution feature offered by the TEE to provide stronger privacy and confidentiality protections for the sensitive transactions. In particular, the sensitive transactions and their data are stored encrypted on the public ledger. When they need to be processed, they are loaded into the protected address space of the TEE, decrypted and consumed therein. The output or updated state of the data and/or transactions are encrypted before being written off the protected address space to the public ledger. Consequently, only ciphertext (i.e., encrypted form) of the sensitive transactions and their data are visible publicly, while their integrity and consistency are accounted for by the TEE and the consensus protocol in use. Thanks to the semantic security of the encryption scheme in use, little, if not none, information can be inferred from the ciphertexts, hence the confidentiality.

\section{Technical Foundations}

\subsection{Trusted Execution Environment}
Intel recently proposed a set of CPUs that are capable of provisioning TEE, in particular Intel SGX. It enables a host to instantiate one or multiple TEEs, or enclaves, simultaneously. An enclave is associated with a CPU-guarded address space which is accessible only by the enclave code; the CPU blocks any non-enclave code’s attempt to access the enclave memory. This effectively isolates the enclave from other enclaves concurrently running on the same host, from the OS, and from other user processes, thereby providing confidentiality and integrity protections for data and code loaded inside the enclaves. Memory pages can be swapped out of the enclave memory, but they are encrypted using the processor’s key prior to leaving the enclave. Enclaves cannot directly execute OS-provided services such as I/O. In order to access those services, enclaves have to employ OCalls (calls executed by the enclave code to transfer the control to non-enclave code) and ECalls (API for untrusted applications to transfer control back to the enclave). These ECalls and OCalls constitute the enclave boundary interface, enabling a communication between the enclave code and the untrusted application to service OS provided functions. The \codename~ leverages Intel SGX to implement the TEEs.

\subsection{Crash Fault Tolerance Consensus Protocol}
A blockchain is essentially a ledger maintained by independent nodes (servers, processors) that are geographically distributed. The consistency of the blockchain relies on these nodes reaching a consensus. A large body of research has been dedicated to consensus protocols, addressing this problem. Consensus protocols primarily tackle either Byzantine failure model or crash failure model~\cite{dang2019autonomous}. In this paper, we pay our attention to a consensus protocol that is designed for the crash failure model~\cite{fast_paxos}. As the name suggests, this threat model assumes that a faulty node crashes (i.e., become indefinitely unresponsive), but never deviates from its intended behaviour.
Our solution builds on the Raft consensus protocol~\cite{raft} for its simplicity. 

The protocol~\cite{raft} is designed for a network of $n$ deterministic nodes. The maximum number of faulty nodes the protocol can tolerate is $f=\frac{n-1}{2}$. 
Each node maintain a log which records an ordered list of transactions. 
The protocol guarantees that logs of non-faulty nodes converge. That is, they record  the same sequence of transactions.
A process can assume one of the following three roles, which are follower, candidate and leader.

Time is split into terms that are numbered with consecutive integers. In each term, there is one node being elected as the leader, while the remaining nodes serve as followers. The leader keeps its authority by  exchanging heartbeat messages with all the followers periodically. Should a follower fail to receive any message from the leader after an election time-out has expired, it considers the leader as having been crashed. It then increments its term value, switches its role to being candidate, and requests for votes to become the new leader. The candidate obtains the leadership if it manages to obtains votes from a majority of other nodes in the network.

In normal operation, the followers respond only to  messages and requests they receive from the leader and candidate, remaining passive otherwise. 
All the transactions (e.g., commands or  requests from the clients) are sent to the leader. It then replicates the transactions on the rest of the network. 
Upon receiving a transaction, the leader insert it as a new entry to its log. The transaction is identified using the leader's current term and an index at which it is inserted to the log. 
Subsequently, the leader broadcasts the entry to all of the followers. 
Upon receiving the entry, the followers inserted it to their logs, and responds the leader with an acknowledgement receipt. 
The leader ensures that the entry has been replicated on a majority of nodes by counting the  acknowledgement it receives. 
Once it receives one from $f+1$ or more nodes, it executes the comment contained  in the entry.
This also commits all other entries preceding the entry in question.
The leader keeps track of the highest index it has committed, and includes such information in subsequent messages it communicates with the followers. This is to inform the later on the committed entries. 
Similar to ~\cite{dang2019autonomous}, by running the Raft consensus protocol inside a TEE that offers attested and isolated execution, \codename~ restricts adversarial behaviours of the faulty nodes, thereby reducing the threat model to crash fault tolerance, to which Raft applies~\cite{mscoco, quorum, 700_bft_protocols}.

\subsection{Scaling Throughput with Parallel Shadow Chains}
The \codename~ comprises multiple parallel shadow chains. Network participants, or nodes or verifiers, in the \codename~ are assigned to a particular shadow chain uniformly at random using a random seed {\tt rnd} generated inside their enclave~\cite{randhound, vrf, pvss1, pvss2}.  Given {\tt rnd}, the nodes
derive their shadow chain assignment by evaluating a random permutation $\pi$ of
$[1:N]$ seeded by {\tt rnd} (with $N$ being the total number of network participants).  $\pi$ is then divided into approximately
equally-sized chunks, each of which represents the verifiers-to-shadow chain assignment.

It is worth emphasizing that there is no upper limit for the number of shadow chains running in parallel. This effectively unleash the transaction throughput of the \codename~, allowing the processing capacity to grow in proportion with the number of verifiers joining the network. 

\vspace{2mm}
\paragraph{Random Seed Generation.} 
The \codename~ exploits TEEs to obtain {\tt rnd} in an efficient manner. The process is partaken by all nodes in the network, thereby assuring the fairness. The Random Seed Generation requires each node to be equipped with a \textproc{RandomnessBeacon} enclave, which is programmed to return fresh, unbiased random numbers subject to a certain probability.
Similar to prior researches~\cite{omniledger, elastico, rapidchain}, this work considers a
synchronous network (i.e., the communication delay $\Delta$ is known a priori) during the distributed
randomness generation procedure.

To obtain its shadow chain assignment, each node in the network invokes its \textproc{RandomnessBeacon} enclave with an epoch number $e$ representing the current period of time.
Given the input $e$, the \textproc{RandomnessBeacon} enclave samples two random
values {\tt q} and {\tt rnd} via two independent invocations of the {\tt
sgx\_read\_rand} function. 
Should $\texttt{q} = 0$, the enclave outputs a signed certificate containing
$\langle e, \texttt{rnd}\rangle$. Otherwise, it returns $\bot$ signalling the node cannot obtain the certificate.
Upon obtaining the certificate, a node broadcast it to the network. After a time $\Delta$,  all nodes in the network should have received all certificates that have been sent. They lock in the lowest {\tt rnd} they receive for epoch $e$, and employs  that value to evaluate its shadow chain assignment~\cite{dang2019towards}.

The security of this procedure is properly analysed in~\cite{dang2019towards}.
The \textproc{RandomnessBeacon} enclave is programmed in such a way that a node can only invoke it once every epoch. This prevents the adversary from selectively discarding the enclave's output so as to bias the final randomness.
In an unlikely event wherein all  nodes fail to receive any message after $\Delta$ (i.e., when no node can obtain  $\langle e,
\texttt{rnd}\rangle$ from its enclave), the nodes increment $e$ and repeat the process. 
The probability of such an event is $P_{\text{repeat}} = (1-2^{-l})^N$ where $l$ is the bit length of {\tt q}. 
This probability can be configured in order to
achieve a desirable trade-off between $P_{\text{repeat}}$ and the communication overhead, which is $O(2^{-l}
N^2)$.  For instance, setting $l=\log(z)$ for some constant $z$, we obtain $P_{\text{repeat}} \approx 0$ and the
communication is $O(N^2)$. Alternatively, if we set $l=\log(N)$, then $P_{\text{repeat}} \approx {\mathrm{e}}^{-1}$ and the communication is $O(N)$.

\vspace{2mm}
\paragraph{Securing individual shadow chain.} 
Nodes on the same shadow chain engage in a consensus protocol to arrive at an agreement on the total order of the transactions incurred on that chain. If a transaction is marked as ``sensitive”, it is processed inside the enclaves with isolated execution, thereby attaining confidentiality protection. 

\vspace{2mm}
\paragraph{Global Order across parallel shadow chains}
We draw the neat idea of establishing one global order for all transactions posted on all shadow chains from ~\cite{ohie}. In particular, transactions in each individual shadow chain are grouped into blocks. Each block is associated with two fields, namely (rank, NextRank), and a chain id, which is the id of the shadow chain it belongs to. These two fields are used to establish total order of transactions across the shadow chains.
 In the total ordering of fully-confirmed blocks, the blocks are ordered by increasing rank values, with tie-breaking based on the chain ids~\cite{ohie}.

Nodes in the network are able to observe all the chains. Thus, they can infer the expected rank of the next block to be recorded on each shadow chain. 
Let us denote by $x$ the largest value among such expected ranks, then $x$ naturally associates with the ``longest'' shadow chain among all the shadow chains. The intuition is that every new block $B$  should help its chain catch up with the current ``longest'' chain. Following this intuition, the node that proposes the new block $B$ should set the NextRank field of $B$ to $x$, or a value greater than $x$. It is also worth emphasizing that $B$'s NextRank should always be larger than $B$'s rank. This constrains is necessary to ensure rank values of blocks on each shadow chains are monotonically increasing. 

Given the (rank, NextRank) fields of blocks are set in the afford mentioned manner, establishing a total order among blocks inhabiting different shadow chains is rather straightforward. For the sake of exposition, let us consider a local view of an honest node at any given time. We denote by  $y_i$ the value contained in the NextRank field of the last partially-confirmed block on the shadow chain $i$, and ConfirmBar  be the minimum among all such values.
It follows  from the setting of (rank, NextRank) that next partially-confirmed block on any shadow chain must have its rank equal to or larger than ConfirmBar value. Consequently, one can consider all partially-confirmed blocks that have rank value smaller than ConfirmBar as fully-confirmed. These fully-confirmed blocks are ordered by their rank values. Should two blocks have equal rank values, tie is broken by their chain ids.

\section{Incentives}
Every blockchain needs a native token. For the sake of exposition, let us call native token of \codename\ by \textit{RT}\footnote{The name of the token in deployment maybe different, and we will announce on the Ritva website once its name has been finalised}.

The RT tokens shall be subject to the hard cap. A portion of the tokens are pre-minted for the development of the \codename. The remaining portion of the  token supply is reserved for rewarding network participants (or verifiers) who contribute their resources to verify the transactions on the \codename. It is worth mentioning that all transactions incurred on the \codename~ network shall be subject to the transaction fee to be collected in RT. A portion of the transaction fee is given to the network participants, while the other is burnt off. This token burn inevitably leads to the depreciation of token supply over time, which translates into an appreciation of the token price against fiat.

\bibliographystyle{plain}
\bibliography{paper}

\end{document}